\documentclass[aps,pra,twocolumn]{revtex4}
\usepackage{amssymb}
\usepackage{graphicx}
\usepackage{subfigure}

\begin{document}

\title{Inelastic collisions in an exactly solvable two-mode Bose-Einstein Condensate}
\author{P. Barberis-Blostein}
\address{Instituto de Ciencias F{\'\i}sicas, Universidad Nacional Aut\'onoma de
M\'exico, Ap. Postal 48-3, 62251 Cuernavaca, Morelos, M\'exico}
\author{I. Fuentes-Schuller}
\thanks{Published before under maiden name Fuentes-Guridi}
\address{Instituto de Ciencias Nucleares, Universidad Nacional Aut\'onoma de M\'exico, A. Postal 70-543, M\'exico D.F. 04510, M\'exico}
\begin{abstract}
Inelastic collisions occur in Bose-Einstein condensates, in some
cases, producing particle loss in the system. Nevertheless, these
processes have not been studied in the case when particles do not
escape the trap. We show that such inelastic processes are relevant
in quantum properties of the system such as the evolution of the
relative population, the self trapping effect and the
probability distribution of particles. Moreover, including inelastic
terms in the model of the two-mode condensate allows for an exact
analytical solution. Using this solution, we show that collisions
favor the generation of entanglement between the modes of the
condensate as long as the collision rate does not exceed the
natural frequency of the system.
\end{abstract}

\maketitle

\section{Introduction}
Bose-Einstein condensates are quantum systems of
great interest since they consist of up to $10^{10}$ particles which
exhibit collective quantum behavior. Recently, multi-component (or
multi-mode) Bose-Einstein condensates have been extensively studied
both theoretically \cite{cirac,internal,theoryspa} and
experimentally \cite{MIT,JILA2,experiment,spatially}. The components
of the condensate are either several spatially separated condensates
\cite{spatially} or several internal degrees of freedom of the
condensed particles \cite{internal}. Unfortunately, the lack of
analytical solutions for these systems limit our understanding of
them. Until now, multi-component condensates have only been studied
through numerical solutions which are always limited by the growing
degrees of freedom of the system.

The simplest multi-component condensate is known as the two-mode
Bose-Einstein condensate. The system consists of condensed particles
in two hyperfine levels \cite{cirac}. Transitions between the levels
are induced via a Josephson-type interaction produced by a laser.
Two body collisions are, together with the laser interaction, the
most relevant physical process in the system. Another incarnation of
the two-mode condensate consists of two spatially separated
condensates coupled through a tunneling barrier corresponding to the
Josepshon-type interaction \cite{theoryspa}. The canonical model
used to describe these systems considers the Josephson-type
interaction and elastic collisions between same and different
particle type \cite{cirac}. The ground state of the model can be
found in the first or second quantization only by numerical means
\cite{cirac,bethe}.

In a recent publication, we introduced a model for the two-mode
Bose-Einstein condensate which includes inelastic collisions
\cite{us}. In our model the inelastic collisions are induced by the
Josephson type-interaction. Including these terms allows for an
exact analytical solution of the model. Several types of inelastic
collisions are known to be relevant in real physical situations
\cite{ine,verification,inelasticexperiment}. The inelastic process
better known in experiments are those which result in particle loss
such as background collisions, three-body recombination,
spin-exchange and dipolar relaxation \cite{verification,tutorial}.
The inelastic processes that we consider here have not been properly
characterized by experiment yet. The main reason for this is that
they do not give rise to particle loss. Nevertheless, there is good
indication that this process are present in real situations. In this
paper, we present a number of theoretical predictions on the
evolution of the relative population of the condensate, the
self trapping effect and on the probability distribution of
particles in the condensates to be corroborated by experiment. We
compare our results with the predictions of the canonical two mode
Bose-Einstein condensate. Experiments could be performed
to compare the predictions of the models. The great advantage of our
model is that it has exact analytical solution which allows for a
complete understanding of the system. Remarkably, we find that the
ground state of the system is the coherent state which was found to
provide a good description of the two-mode Bose-Einstein condensate
\cite{coherent,choi}. We take advantage of the simplicity of the
model to show under what conditions maximally entangled states
between the modes can be created and the role of collisions in the
generation of entanglement.


\section{Particle loss due to inelastic collisions in Bose-Einstein condensates}
Inelastic collision in Bose-Eistein condensates occur with high
frequency and have been studied extensively
\cite{tutorial,fesh,verification,ine}. These processes are well
known since they commonly result in particle loss. For this reason
there has been considerable efforts to suppress them in experimental
situations \cite{tutorial,verification}. The main inelastic
processes known are background collisions, three-body recombination,
spin exchange and dipole relaxation. Atoms in a condensate are
commonly knocked out of the trap by collisions with untrapped,
room-temperature molecules in the background chamber. In the three
body recombination process three particles collide and form a
molecule which is no longer trapped by the potential. Particles are
also lost when, during a collision, the atoms make transitions to
different energy levels which are not trapped by the potential.
Spin-exchange and dipolar relaxation are important mechanism of
particle loss which are closely related to the process considered in
our model. After a collision either or both atoms exit in a
different spin state. If the spin re-orientation energy is larger
than the depth of the trapping potential this mechanism might result
in the loss of the particles from the trap \cite{fesh}. In dipolar
relaxation one or two particles in the higher hyperfine level decay
to the lower level during the collision due to their dipole-dipole
interaction. The excess of energy is then transferred to an excess
of momentum which produces the particle to escape the trap. In our
model the inelastic process occurs but the excess of energy is not
enough to help the particle escape from the trap. Theoretically,
inelastic processes are included in the studies of the condensates
by considering classical rate equations for the particle loss
\cite{verification}. But in the literature there is a lack of
experimental and theoretical analysis on inelastic process which do
not give rise to particle loss. Our model exemplifies the importance
that these processes might have in the behavior of the system.

\section{Two-mode Bose-Einstien condensate with inelastic collisions}

Our model of the two mode Bose-Einstein condensate consists of the
following Hamiltonian,

\begin{eqnarray}\label{eq:hbec}
H_{2}&=&A_{0}+\delta\omega(a^{\dagger}a-b^{\dagger}b)\\
&+&\lambda(e^{i\phi}a^{\dagger}b+e^{-i\phi}ab^{\dagger})+\mathcal{U}\, a^{\dagger}b^{\dagger}a b\nonumber\\
&+&\Lambda(e^{2i\phi}a^{\dagger}a^{\dagger}bb+h.c.)\nonumber\\
  &+&\mu((a^{\dagger}a^{\dagger}ab-b^{\dagger}a^{\dagger}bb)e^{i\phi}
  +h.c.).\nonumber
 \end{eqnarray}

The modes $a^{\dagger},a$ and $b^{\dagger},b$ with frequency
difference $\delta\omega$, correspond to either atoms with two
different hyperfine levels \cite{JILA2} or alternatively, two
spatially separated condensates \cite{MIT}. The Josephson-type
interactions is induced by applying a laser \cite{JILA2} or a
magnetic field gradient \cite{MIT}. In our Hamiltonian the
Josephson-type term, in which one particle is annihilated in one
mode and created in the other, has coupling constant $\lambda$ and
phase $\phi$. The terms with four bosonic operators describe
two-particle elastic and inelastic collisions. The elastic
collisions have interaction strength $\mathcal{U}$. The inelastic
collisions have interaction strength $\mu$ when two particles in the
same mode collide and one of them is transformed into the other and
interaction strength $\Lambda$ when the collision transforms two
particles in one mode into the other. Ignoring the inelastic terms
in Eq.(\ref{eq:hbec}), we find that our model coincides with the
canonical Josephson Hamiltonian \cite{cirac} when the rate of
collisions of same particle type is equal. The assumption of equal
collision rates for the same particle type is also made in
\cite{cirac} in order to find approximate and numerical solutions. Our
model has the same number of free parameters as the canonical
two-mode Hamiltonian. The only difference is that the Hamiltonian in
Eq.(\ref{eq:hbec}) includes inelastic collisions, which are usually
present in real BECs \cite{ine}.

Including the correct rate of inelastic collisions in the
Hamiltonian allows for an analytical solution. The solution is
simply found by considering the following Hamiltonian
$H_{0}=A_{1}(a^{\dagger}a-b^{\dagger}b)+A_{2}(a^{\dagger}a-b^{\dagger}b)^{2}$
with eigenstates $|N,m\rangle$ where $N$ is the eigenvalue of the
total number operator
$\hat{N}=n_{a}+n_{b}=a^{\dagger}a+b^{\dagger}b$ and $m$ the
eigenvalue of the relative population
$\hat{m}=a^{\dagger}a-b^{\dagger}b $. Since the number of particles
in the system $N$ is constant, $m$ is restricted to values
$m=-N,...,N$. $H_{0}$ describes a two-mode condensate with no
Josephson-type interaction, i.e, no laser interaction or tunneling
barrier. The parameter $A_{2}$ corresponds the two-body elastic
scattering probability and $A_{1}$ to the energy difference between
the modes. It is easy to verify that if the parameters of the model
satisfy the following conditions
\begin{eqnarray}\label{eq:coef}
A_{0}&=&A_{2}(\cos^{2}{\theta}N^{2}+\sin^{2}{\theta}N)/4\, ,\nonumber\\
\delta\omega&=&(A_{1}\cos{\theta})/2\, ,\nonumber\\
\lambda&=&(A_{1}\sin{\theta})/2\, \nonumber\\
\mathcal{U}&=&A_{2}(1-3\cos^{2}{\theta})/4\, ,\nonumber\\
\mu&=&(A_{2}\cos{\theta}\sin{\theta})/2\, ,\nonumber\\
\Lambda&=&(A_{2}\sin^{2}{\theta})/4.
\end{eqnarray}
then the Hamiltonian can be written as $H_{2}=UH_{0}U^{\dagger}$
where where $U$ is the two-mode displacement operator $U=e^{\xi
a^{\dagger}b+\xi^{\ast}ab^{\dagger}}$ with displacement parameter
$\xi=\theta e^{i\phi}$. The solution of the Hamiltonian
(\ref{eq:hbec}) is simply $U^{\dagger}|N,m\rangle$ with energy
$\mathcal{E}_{m}=A_1 m+A_2 m^2$.

 Note that by fixing the free parameters
$\delta\omega$, $\lambda$ and $U$ the inelastic collision constants,
$\Lambda$ and $\mu$, are determined. This is because in our model,
the inelastic collisions are produced by the effect of the
Josephson-type interaction on colliding particles. This physical
relationship is mathematically expressed by the relationship between
the coefficients. The relationship between the parameters imply
that, for fixed $\theta$, the scattering probabilities for inelastic
collisions are higher if the scattering for elastic collisions are
more probable. If $\sin(\theta)=0$, the Josephson coupling is zero
and there is no inelastic processes occurring in the system. There
are only elastic collisions between particles of the same type. We
also see that for a resonant laser or a symmetric well
$\delta\omega=0$, which corresponds to $\cos\theta=0$, there are
elastic and inelastic collisions but the inelastic interaction
always involves the exchange of two particles simultaneously.
We would like to emphasize that if for a specific physical system,
the rates of inelastic to elastic collisions does not hold or cannot
be arranged by external manipulation, an analytical solution cannot
be found using our method for such a condensate. Fortunately, in the
laboratory the rate of elastic and inelastic collisions can be
manipulated, for example, by applying a magnetic potential
\cite{roberts} making possible to meet the experimental values for
the production ratios of those terms. The advantage of having a
region in the parameter space of the system where we can find an
exact solution is that it is then possible to find solutions for
other parameters using perturbation theory.

Due to the simplicity of our solution the ground state
$U^{\dagger}|N,m_0\rangle$ of $H_{2}$ is trivially found by
minimizing the energy $E^{2}_m=A_1 m+A_2 m^2$ with respect to m. For
$A_2>0$, $m_0$ is the nearest integer to $-A_1/(2 A_2)$ or $m_0=-A_1
N/|A_1|$ when $|-A_1/(2 A_2)|>N$. For $A_2<0$, the minimum
corresponds to $m_0=N$ if $A_1<0$ or $m_0=-N$ otherwise. The
solution for $A_2<0$ and $A_1>0$ is the spin coherent state which
has been argued to describe the Bose-Einstein condensate well enough
\cite{choi,coherent}.

\section{Inelastic collisions in a double well} In the physical
realization of the model, where the two modes $a,a^{\dagger}$ and
$b,b^{\dagger}$ of the condensate correspond to particles in two
spatially separated condensates, A and B, the Josephson-type
interaction corresponds to the tunneling of particles through a
barrier with probability proportional to $\lambda$.

The canonical model, as well as our model, considers elastic
collisions between the particles in each well given by the terms
proportional to $a^{\dagger}aa^{\dagger}a$ and
$b^{\dagger}bb^{\dagger}b$. Two particles in one well are
annihilated and two particles in the same well are created. Note
that in our model these terms are part of the constant term $N^2$
which is included in the parameter $A_0$. Both models also consider
collisions between particles in different wells given by
$a^{\dagger}ab^{\dagger}b$. One particle in well A and one particle
in well B are annihilated and one particle in well A and one
particle in well B are created. These collisions occur in the region
where the spatial wave functions of each well overlap.
In all these elastic
interactions the particle number in each well is conserved. The new
feature in our model is that we consider collisions occurring in the
overlapping region in which particles in a well end up in the other
well after the collision. We consider terms in which two particles
in one well collide (close to the tunneling barrier) and one or two
of them are simultaneously tunneled to the second well. We also
include situations in which particles from different wells collide
in the overlapping region and both of them end up in the same well
after the collision. The probability of such inelastic events
depends on the overlapping region between the condensates, which is a function
of the tunneling probability and the
trapping potential. The probability of inelastic events is also a
function of the probability of elastic events which depend on the
scattering lengths of particles in the system. It seams logical and
natural to consider such inelastic processes. Note that there are
other theoretical motivations to support the inclusion of such terms
\cite{indus}. The specific relationship between the parameters for
which the model has an exact analytical solution corresponds to a
special type of potential.

\section{Inelastic collisions between particles in the presence of
a laser field}

In the physical realization of the model where the two modes
correspond to two hyperfine levels, the Josephson-type interaction
corresponds to the interaction of a laser field which induces
transitions between the atoms hyperfine levels. In these picture,
same and different spice particle can collide. The inelastic
collisions correspond to the collision of particles in the presence
of a laser field. Energy is absorbed and liberated into the field
during collision in which the particles change their hyperfine
levels. Although a detailed microscopic calculation of this process
has not been yet calculated, it has been observed in experiments
that the rate of inelastic collisions between atoms is increased
when there is an interaction with a laser field
\cite{inelasticexperiment}. We provide in this paper several
theoretical predictions to be experimentally tested of the effects
of inelastic collusion which do not give rise to particle loss in
the condensate.
\section{Effects of inelastic collisions in the system}
In this section, we analyze the effects of the inelastic collision
terms added to the Hamiltonian in the behavior of the two-mode
Bose-Einstein condensate. We study the particle distribution of the
ground state and the evolution of relative population in each mode.
We will show that the presence of inelastic collisions have
important consequences in the self trapping effect.

\subsection{Particle distribution of the ground state}

\begin{figure}[!ht]
\subfigure{
\includegraphics[width=1.5in]{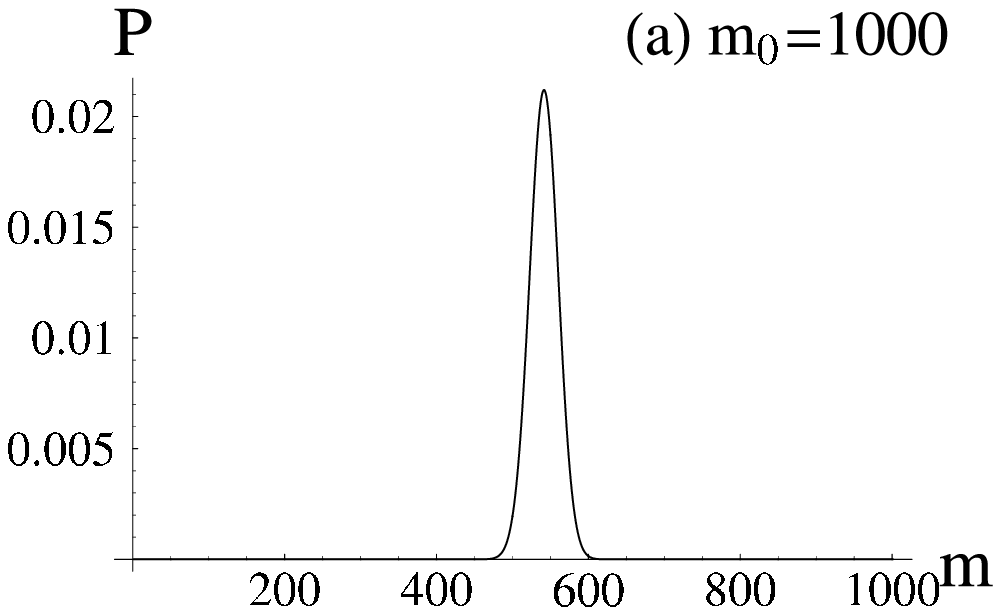}\label{fig:ma}}
\subfigure{\includegraphics[width=1.5in]{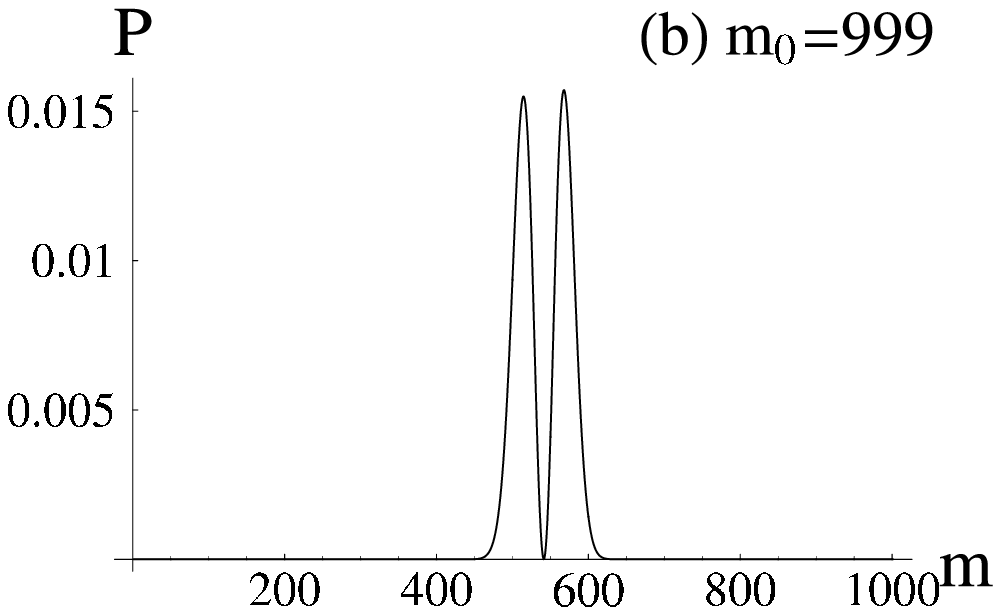}\label{fig:mb}}\\
\subfigure{
\includegraphics[width=1.5in]{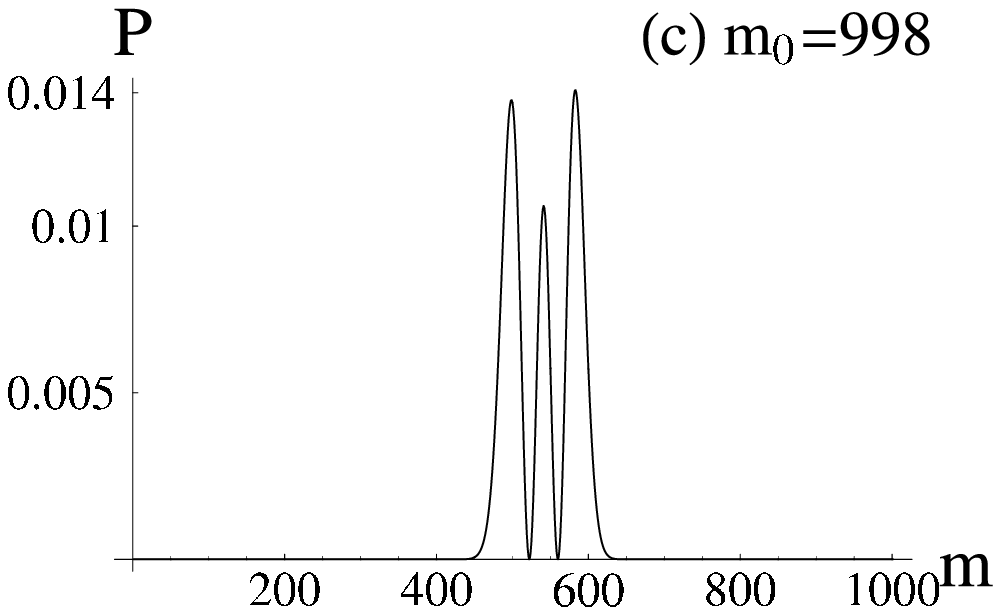}\label{fig:mc}}
\subfigure{
\includegraphics[width=1.5in]{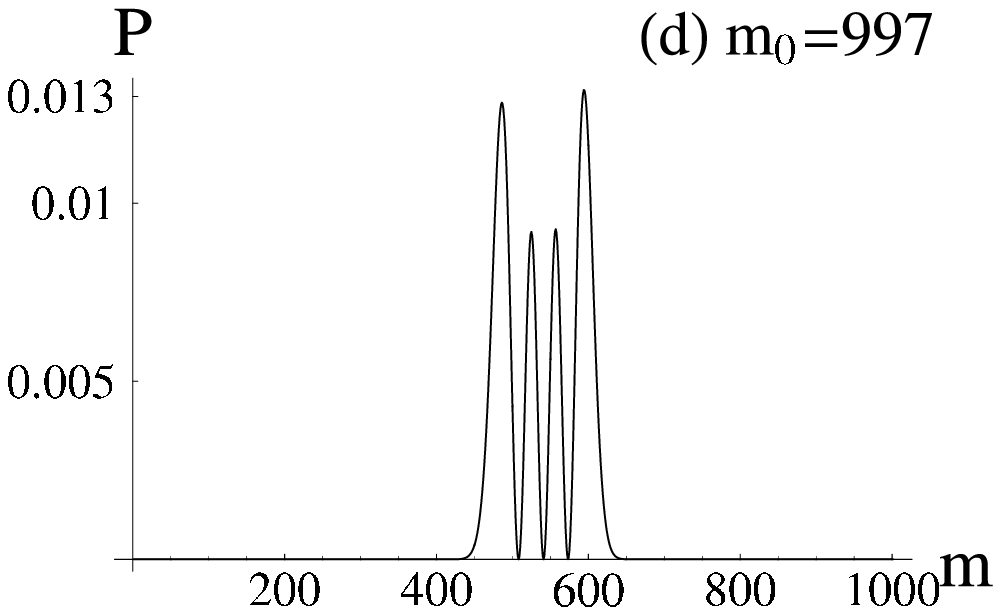}\label{fig:md}}
\caption{\label{fig:m} Ground state relative population distribution
for $1000$ atoms and $\theta=1$. Different $m_0$
  correspond to a different rate between the parameters $A_{1}$ and $A_{2}$. Quantum
  superposition appears when $m_0<1000$.}
\end{figure}

Let us first consider the analytical solution to our model. The
particle distribution $P=|\langle N,m|\psi_0\rangle|^2$ of the
ground state $|\psi_0\rangle=U^\dagger|N,m_0\rangle$  is equal to
the modulus square of the Clebsch-Gordon coefficients (Wigner)
$P=|d_{m,m_{0}}^{N}|^2$ where,
\begin{widetext}
\begin{equation}
d_{m,m_{0}}^{N}=\sum_k(-1)^{k-m_0+m}\frac{\sqrt{(N+m_0)!(N-m_0)!(N+m)!(N-m)!}}{(N+m_0-k)!k!(N-k-m)!(k-m_0+m)!}
(\cos(\theta/2))^{2N-2k+m_0-m}(\sin(\theta/2))^{2k-m_0+m}.
\end{equation}
\end{widetext}
The sum must be done over k whenever none of the arguments
of factorials in the denominator are negative. Different ground
states parameterized by $m_0$ are obtained by changing the rate
$A_1/A_2$ since $m_{0}$ is the closest integer to $-A_{1}/2A_{2}$
(for $|-A_{1}/2A_{2}|<N$ otherwise, $m_0=N$). When $m_0=N$, the
particle distribution is
\begin{equation}
|d_{N,m}^{N}|^{2}=\sqrt{\frac{2N!}{(N+m)!(N-m)}}\cos(\theta/2)^{N+m}\sin(\theta/2)^{N-m}
\end{equation}
which corresponds to a distribution with one maximum.
We plot an example for $N=1000$ particles and $\theta=1$ in Fig.
\ref{fig:ma}. As we change the rate $A_1/2A_2$ from 1000 to 977, the
distribution changes abruptly from a distribution with a single
maximum to a distribution with several peaks corresponding to
multiple cat states as shown in Fig. \ref{fig:m}. These
discontinuous changes occur when the system undergoes first order
transitions corresponding to level crossings. The structure of the
ground state changes abruptly and this is reflected in the
probability distribution. As $m_{0}$ approaches zero, the number of
maxima in the distributions grows. Controlling the inelastic
parameters allows us to prepare the condensate in such superposition
states.

\begin{figure}[!ht]
\subfigure{
\includegraphics[width=1.5in]{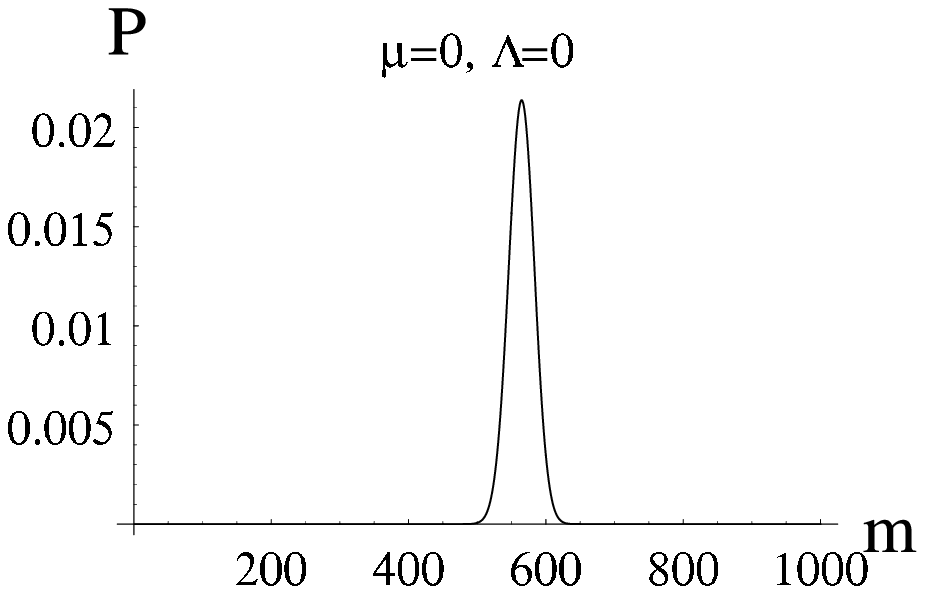}}
\subfigure{\includegraphics[width=1.5in]{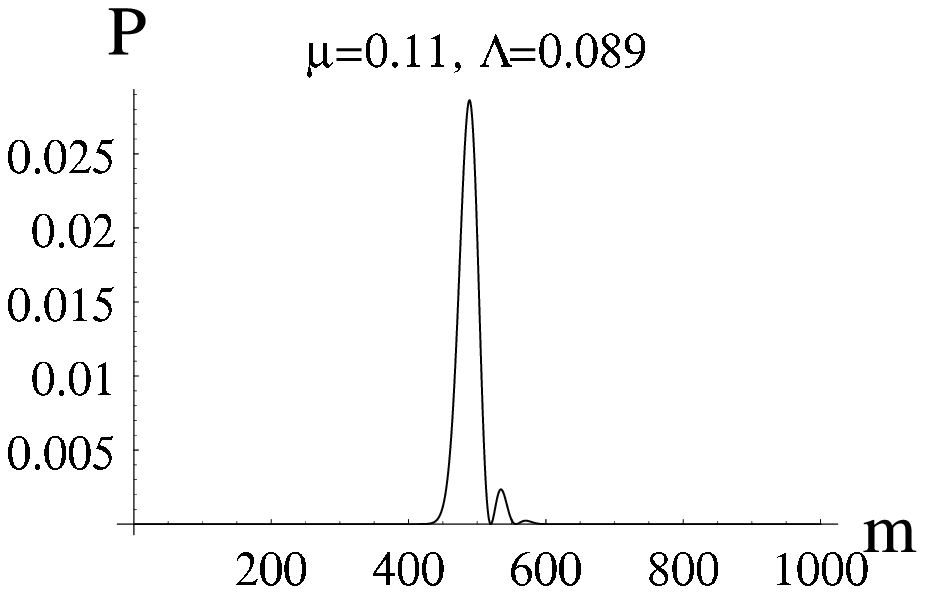}}\\
\subfigure{
\includegraphics[width=1.5in]{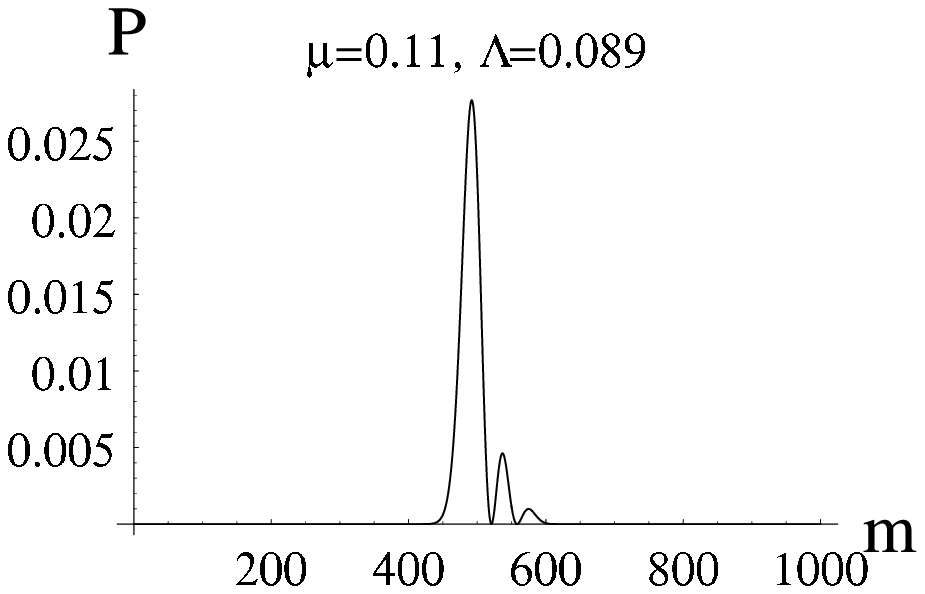}}
\subfigure{
\includegraphics[width=1.5in]{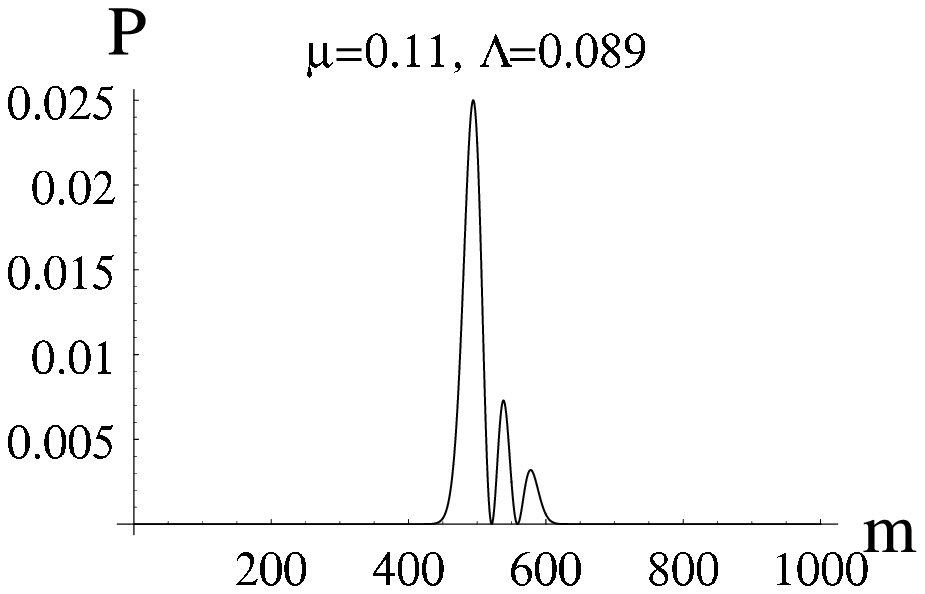}}\\
\subfigure{
\includegraphics[width=1.5in]{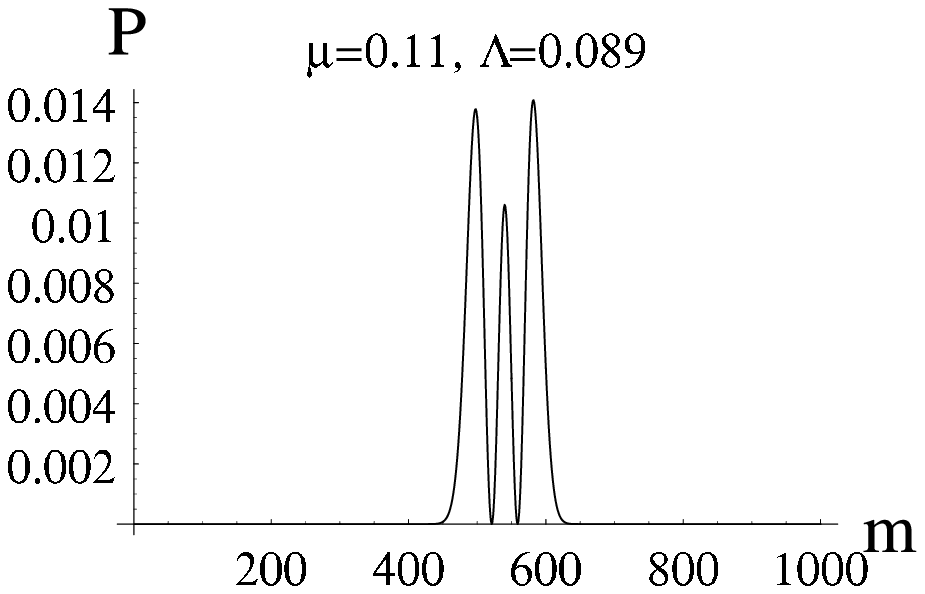}}
\subfigure{
\includegraphics[width=1.5in]{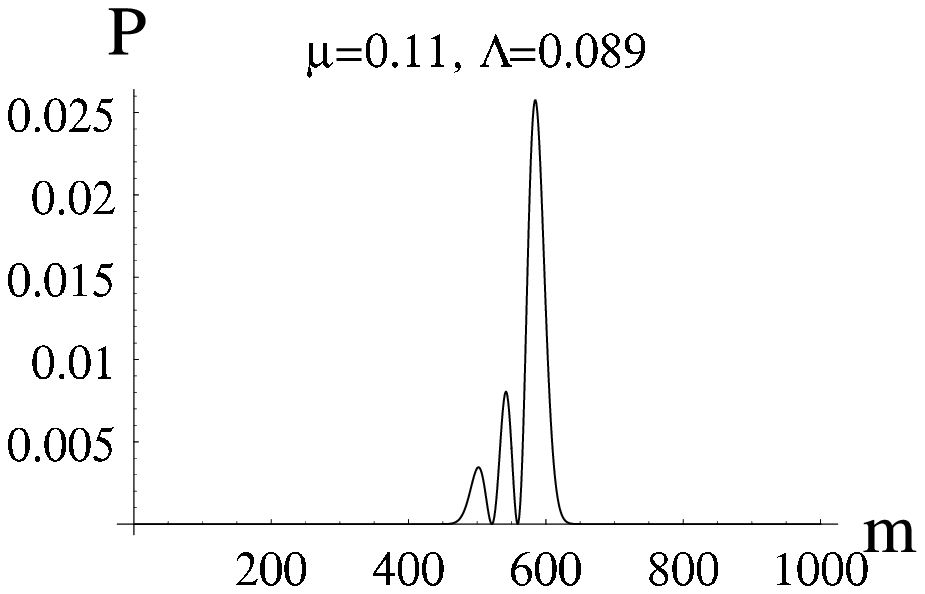}}
\caption{\label{fig:mcont} Ground state relative population
distribution for $1000$ atoms. $\delta\omega=109J$, $\lambda=487J$ and
$\mathcal{U}=0.214027J$.}
\end{figure}

Now consider the Hamiltonian (\ref{eq:hbec}) with general
parameters. In this general case it is not possible to find an
analytical solution. Nevertheless we find, by numerical means, the
probability distribution of particles in the ground state. In Fig.
\ref{fig:mcont} we show the probability distribution for $N=1000$.
We leave $\delta\omega=109J$, $\lambda=487J$ and
$\mathcal{U}=0.214027J$ fixed and vary $\mu$ and $\Lambda$ as shown
in the figure. By varying these parameters we show that the
transition from different probability distributions of the ground
state can be done smoothly. The change from a distribution
with one maxima to a superposition distribution is no
longer abrupt. We find the abrupt transition only when the
parameters change along the values corresponding to those for which
the system has exact analytical solution. In the canonical model,
which does not include inelastic collisions, the change from a
single peaked distribution to a double superposition is
also smooth as shown in \cite{cirac}. The main difference observed
between the canonical model and our model is that in the canonical
model only single and double superposition distributions
are observed. In our model, it is possible to obtain the
superposition of N distributions. Therefore, the presence of more
than two components in the probability distribution is a clear
footprint of the presence of inelastic collisions in the system.

\subsection{Inelastic collisions in the evolution of the relative population and the self trapping effect}
\begin{figure}[h!]
\subfigure[]{\label{fig:jzinea}
\includegraphics[width=2.3in]{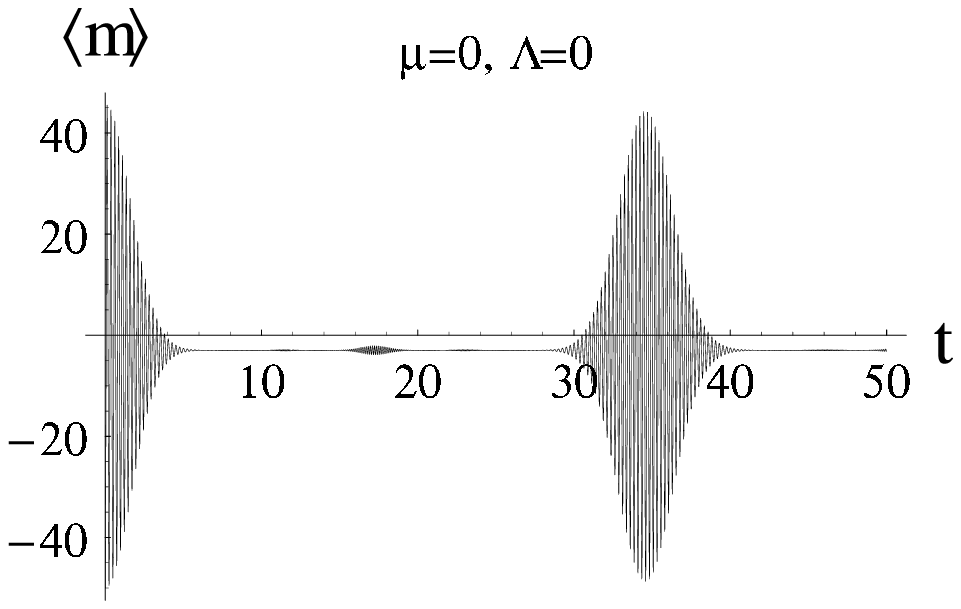}}\\
\subfigure[]{\label{fig:jzineb}
\includegraphics[width=2.3in]{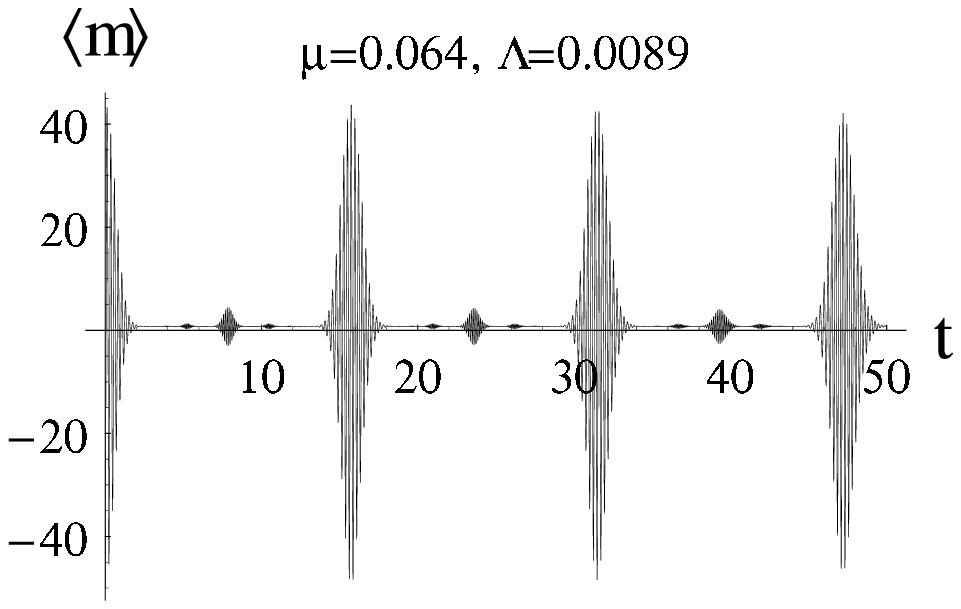}}\\
\subfigure[]{\label{fig:jzinec}
\includegraphics[width=2.3in]{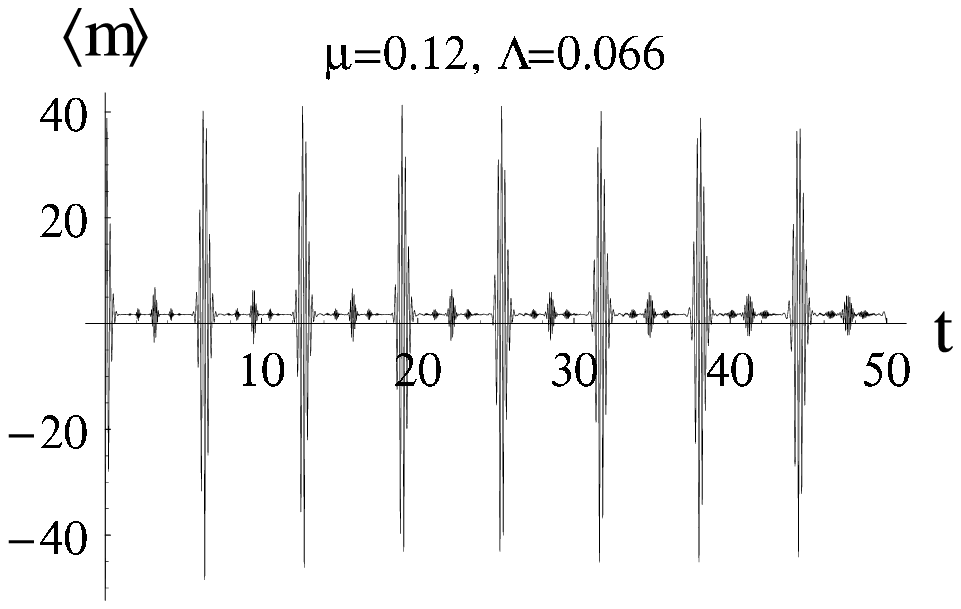}}\\
\subfigure[]{\label{fig:jzined}
\includegraphics[width=2.3in]{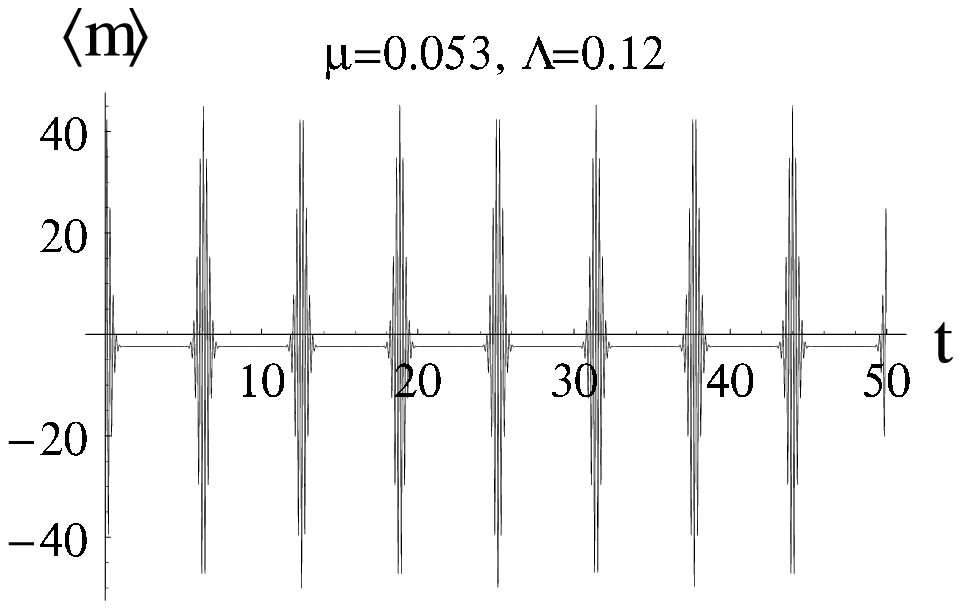}}
\caption{\label{fig:jzine} Evolution of the relative population
$\langle m\rangle$,$\delta\omega=109J$, $\lambda=487J$ and
$\mathcal{U}=0.214027J$ and vary $\mu$ and $\Lambda$ are shown in the
figure.}
\end{figure}

We now calculate the evolution of the average relative population
$\langle a^\dagger a-b^\dagger b\rangle$, for the initial condition
$|\psi(t=0)\rangle$. In the case where the system has an analytical
solution, the evolution of the relative population for a given
initial state $|\psi(t=0)\rangle=\sum_{m=-N}^{N} C_m
U^\dagger|N,m\rangle$, with coefficients $C_{m}$, is given by
\begin{eqnarray}\label{eq:jzev}
\langle a^\dagger a&-&b^\dagger
b\rangle=\cos\theta\sum_{-N}^{N}m\,|C_m|^2\\&-&\sin\theta\sum_{-N+1}^{N} C_{m}
C_{m-1}(N(N+1)-m(m-1))^{1/2} L_{m}\nonumber\\
L_{m}&=&\cos(\phi+(E_{m-1}-E_m)\,t)
\end{eqnarray}
In the general case, the evolution of the relative population is
calculated numerically. In Fig. (\ref{fig:jzine}) we plot the
evolution of the relative population for $\delta\omega=109J$,
$\lambda=487J$ and $\mathcal{U}=0.214027J$ and vary $\mu$ and
$\Lambda$ as shown in the figure. Fig\ref{fig:jzinea} correspond to
the evolution of the canonical model (without inelastic collisions).
In Figs.(\ref{fig:jzineb}-\ref{fig:jzinec}) we plot the evolution of
the general model which includes inelastic terms. We considered
several cases with different amount of inelastic collisions. Until
here all the results where numerical. Finally, the exact analytical
solution given by Eq. (\ref{eq:jzev}) is plotted in
Fig.\ref{fig:jzined}. In the canonical model the collapse and
revivals are not uniform. Each revival becomes wider and at some
point one revival gets mixed up with the next one. We can see that
adding inelastic collisions changes drastically the behavior of the
system. As the rate of elastic and inelastic collision approaches
the values for which we have analytical solutions, we observe that
the revivals start becoming more uniform. We can go from the
solution of the canonical model to the analytical solution of our
model continuously by varying the inelastic collision parameters.
Finally, for the analytical solution of the model, obtained when the
rate of elastic to inelastic collisions is equal to $1/2$, the
evolution of the relative population is periodic and all revivals
have the same width. One can see in Fig.\ref{fig:jzined} that the
structure of the revivals repeats after two collapses.

We take advantage of having an analytical solution to calculate
(using Eq. (\ref{eq:jzev})) the time at which the oscillations of
the relative population collapse and the period at which the
collapse and revivals occur. In order to calculate the time of
collapse we find the condition when the different terms in Eq.
(\ref{eq:jzev}) cancel each other. This occurs when the phase
difference between the cosines in $L_m$ is $(2n+1)\pi$, where $n$ is
an integer. Therefore, we obtain the condition
$(E_{m-1}-E_m)t-(E_m-E_{m+1})t=\pi (2n+1)$ which implies that the
time of collapse is $t_r=(2n+1)\pi/(2 A_2)$. We will assume that
$t=0$ corresponds to exactly half of the revival (as in Fig.
\ref{fig:jzine}). Therefore, $t_r(n=0)$ corresponds to half the
distance between the revivals.

\begin{figure*}
\subfigure[]{
\includegraphics[width=8in]{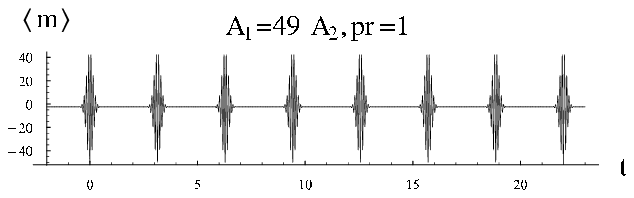}}\\
\subfigure[]{
\includegraphics[width=8in]{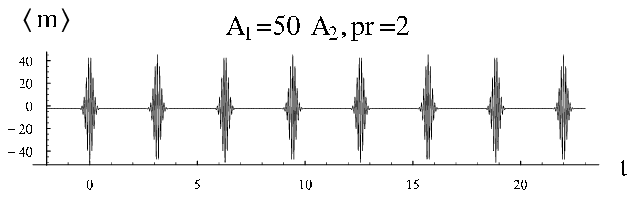}}\\
\subfigure[]{
\includegraphics[width=8in]{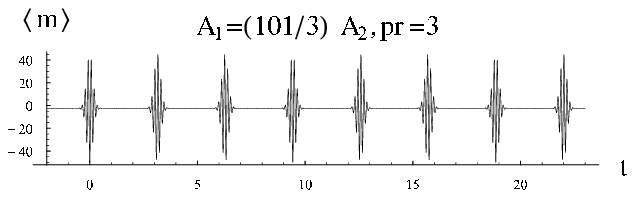}}\\
\subfigure[]{
\includegraphics[width=8in]{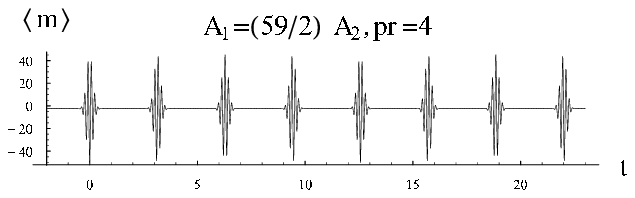}}\\
\caption{\label{fig:jzpr} The evolution of the relative population
$\langle m\rangle$ is plotted for the following values $theta=1.35$,
$A2=1$ and $A1=49,50,101/3,59/2$. The values are chosen so
$p_r=1,2,3,4$}
\end{figure*}

To calculate the period of the evolution of the relative population
we suppose that the initial time is $t_0=0$ and consider a later
time $t_1>t_0$ at which the average relative population is the same
$\langle a^{\dagger}a-b^{\dagger}b\rangle(t_0)=\langle
a^{\dagger}a-a^{\dagger}a\rangle(t_1)$. If for every integer $m$
exists an integer $n_m$ where the following condition is satisfied,
\begin{equation}\label{eq:t1}
(E_{m-1}-E_m)t_1=2\pi n_m\, ,
\end{equation}
from Eq. (\ref{eq:jzev}) we can see that the evolution of all the
terms $L_m$ is identical from $t_1$ as it is from $t_0$. If we consider $t_1$
to be the smallest number which satisfies Eq. (\ref{eq:t1}) the
function $\langle a^{\dagger}a-b^{\dagger}b\rangle$ is periodic with
period $t_1$. We can write Eq. (\ref{eq:t1}) as,
\begin{equation}
(-A_1-A_2 (2 m-1)) t_1=2 \pi n_m\ \label{eq:cond1}\, .
\end{equation}
If $A_1/A_2$ is not a rational number Eq. (\ref{eq:cond1}) is never
satisfied and the function is not periodic. We will have collapse
and revivals, but the particular form of every revival would be
different. When $A_1/A_2$ is a rational number, the time $t_1$
corresponds to the period, and $pr=t_1/(2 t_r(n=0))$ is the period
of the relative population in units of number of revivals. Remember
that $t_r(n=0)$ corresponds to half the distances between
 revivals. In other words, $p_r=k$ means that after $k$ revivals
the particular structure of the first revival repeats itself.
Considering $A_1=c A_2$ with $c=p/q$ and $p,q$ integers, one can se
that if $-q+p$ is even then $pr=q$, and if $-q+p$ is odd then
$pr=2q$. An example is shown in Fig.\ref{fig:jzpr}, where the
evolution of the relative population is plotted for the following
values $\theta=1.35$, $A2=1$ and $A1=49,50,101/3,59/2$. The values
are chosen so that $p_r=1,2,3,4$.

In the canonical model ($\mu=0,\Lambda=0$), there are no Rabi
oscillations if $\mathcal{U}\gg \lambda$. The condensate remains
trapped in one mode if the scattering probability is much larger
than the Josephson-type interaction. This corresponds to the
well-known self trapping effect \cite{millburn}. Now consider our
model, when the parameters satisfy conditions (\ref{eq:coef}) and
the evolution of the relative population is given by Eq.
(\ref{eq:jzev}). If $A_2$ and $A_1$ are fixed, the ratio
$\mathcal{U}/\lambda$ is maximum for $\theta=0$. Under these
circumstances, it can be trivially seen from Eq. (\ref{eq:jzev})
that there are no oscillations and thus, we observe self trapping.
This case is trivial since there is no Josephson-type interaction.
However, if we consider $\theta$ fixed, we find that the inelastic
collisions can prevent self trapping from occurring even if
$\mathcal{U}\gg\lambda$. For instance, consider $\theta=\pi/2$,
there is a collapse at time $t_r$ (defined above). In this case,
although $A_2\gg A_1$ (implying that $\mathcal{U}\gg \lambda$) there
is never self trapping.

\section{Entanglement}
Entanglement is a property of multipartite quantum states that
arises from the tensor product structure of the Hilbert space and
the superposition principle. It plays an important role in the
understanding of many body quantum systems. Moreover, it is considered to be a resource for quantum information tasks %
\cite{nielsen00} such as teleportation and has applications in
quantum control and quantum simulations \cite{control}. It is
therefore of interest to find the necessary conditions to prepare
states with high degrees entanglement.

Entanglement between the modes of a two component Bose-Einstein
condensate have been studied in the canonical model
\cite{entanglement,choi}. In this section, we calculate entanglement
between the modes using the analytical solution to our model and
show the conditions to generate maximal entanglement in the system.
We analyze in detail the role of collisions in the generation of
entanglement.

Quantifying entanglement for pure bipartite systems is simple.
Entanglement between the systems is given by the von-Neumann entropy
$S(\rho_{a})=-tr(\rho_{a}\log_{2}(\rho_{a}))$, where $\rho_{a}$ is
the reduced density matrix. The entanglement between the modes for
the ground state of our system $U^{\dagger}|N,m_{0} \rangle$ is
given simply by $S(\theta, N,
m_{0})=-\sum{m}|d_{m_{0},m}^{N}(\theta)|^{2}\log_{2}|d_{m_{0},m}^{N}(\theta)|^{2}$
where $d_{m_{0},m}^{N}(\theta)$ are the Clebsch-Gordon coefficients.
The entanglement is a function of the angle $\theta$, the number of
particles $N$ and $m_{0}$. The ground state is determined by the
parameter $m_{0}$ which depends on the rate $A_{1}/2A_{2}$. This
parameter regulates the size of the elastic collision probability in
comparison to the difference of frequency of the modes in the
absence of Josephson-type interaction.

It is easy to see from the formula that the entanglement of the
state $m_{0}=N$ is maximum when $\theta=\pi/2$ independently of the
number of particles in the condensate. This means that when the
collisions rate is much smaller than the frequency difference
between the modes in the absence of Josephson-type interaction
$A_2\ll A_1$, the entanglement generated after turning on the
interaction, will be maximum if the double well is symmetric with a
high tunneling probability. In a two species Bose-Einstein
condensate, maximum degrees of entanglement will be generated, in
this case, by turning on an intense resonant laser. Note that when
$m_{0}=N$ the probability distribution has a single component.
\begin{figure}[h!]
\subfigure[]{
\includegraphics[width=2in]{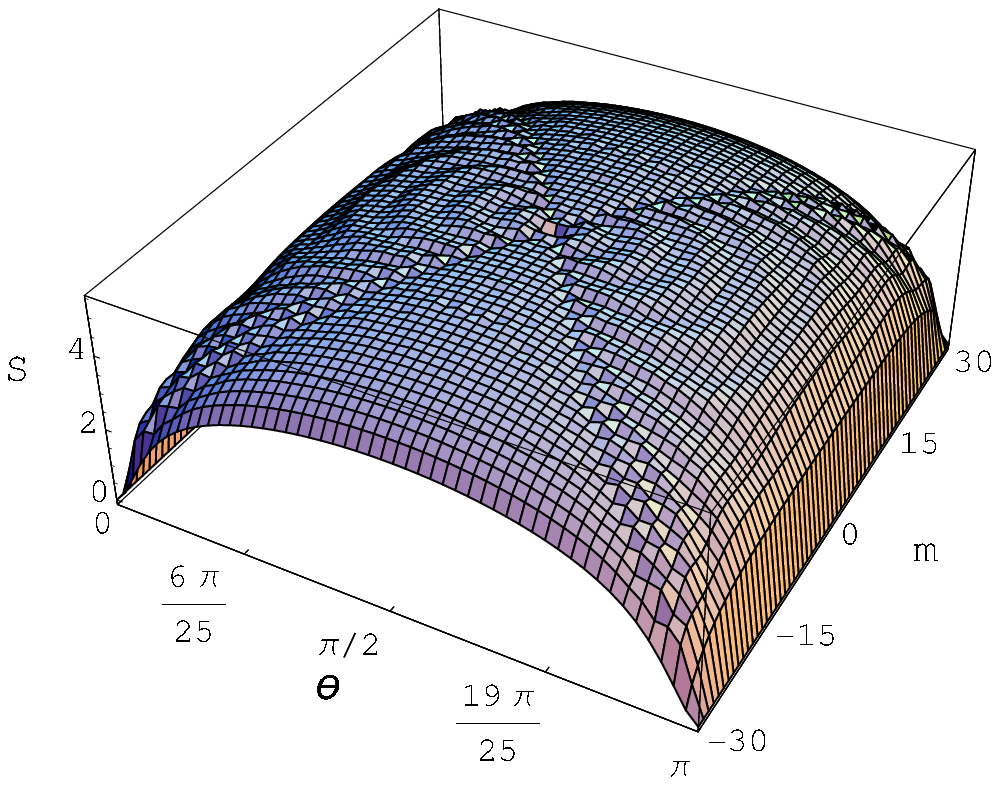}}
\subfigure[]{\includegraphics[width=2in]{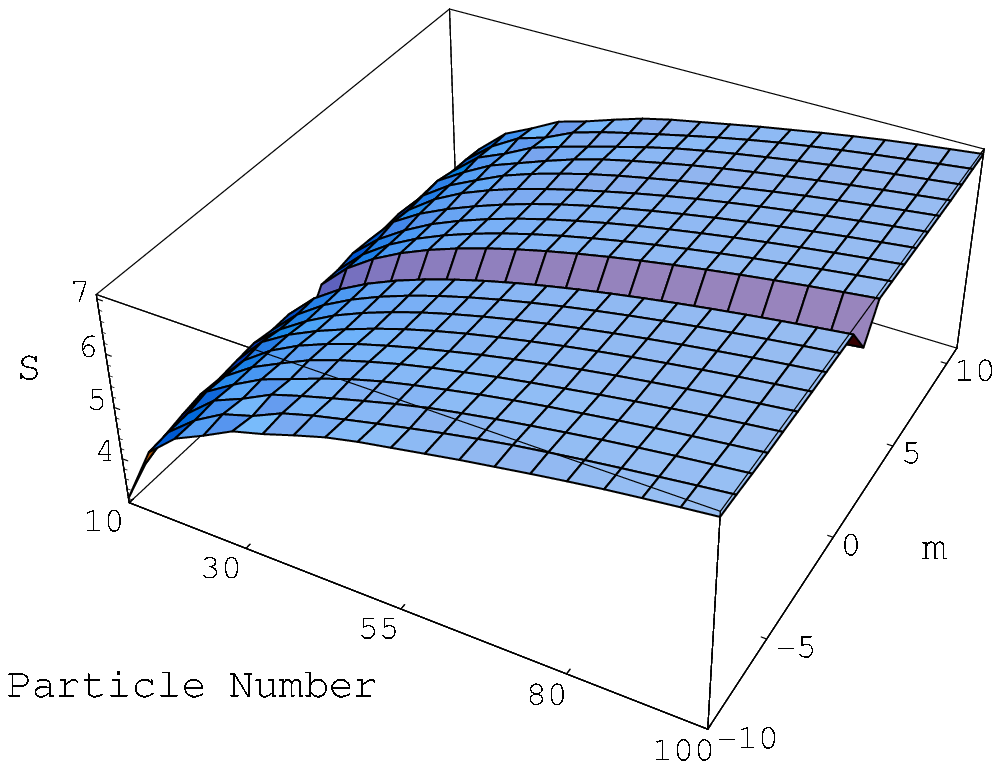}}
\caption{\label{fg:ent}Entanglement
a) as a funtion of $\theta$ and $m_{0}$ for $N=100$ and b) as a
funtion of $N$ and $m_{0}=0$ for $\theta=\pi/2$.}
\end{figure}

\begin{figure}[h!]
\subfigure[]{
\includegraphics[width=2in]{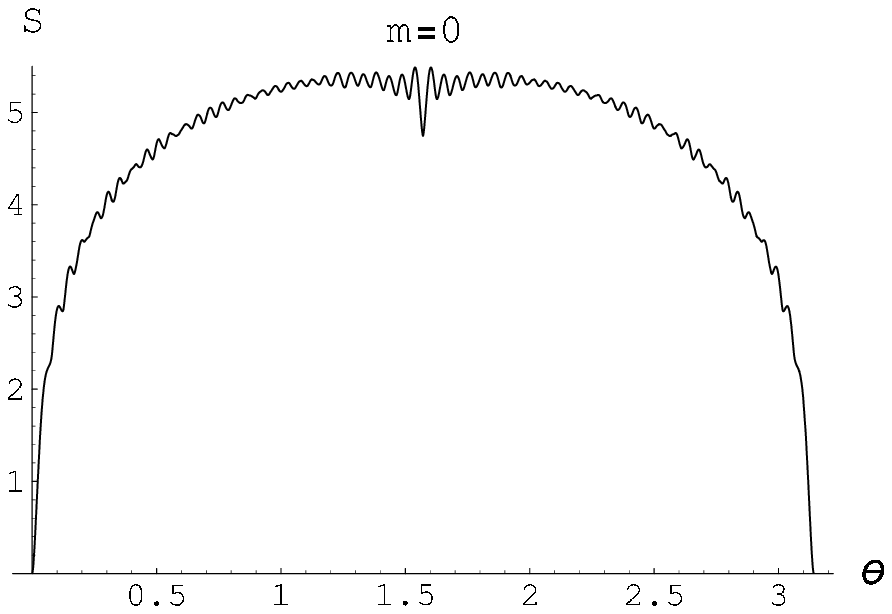}}
\subfigure[]{
\includegraphics[width=2in]{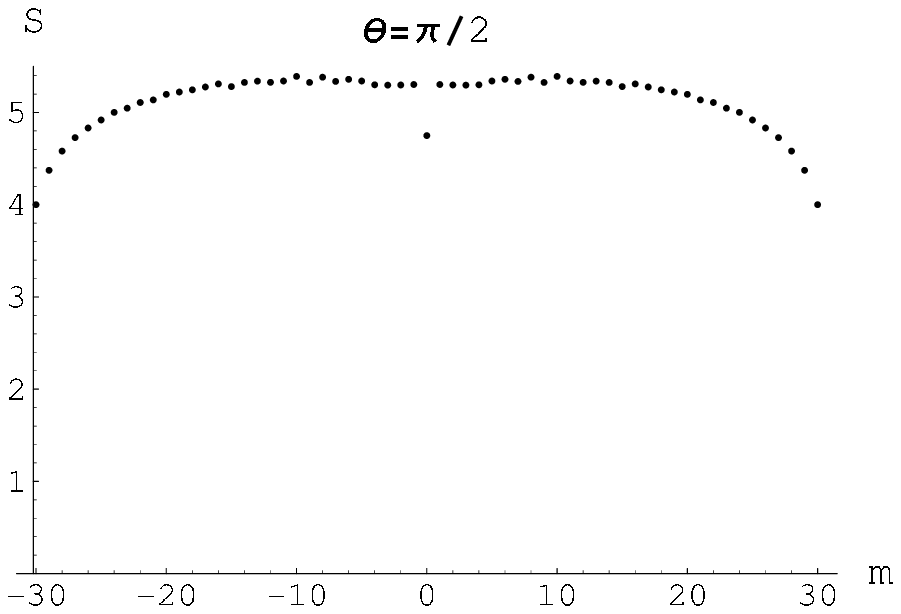}}
\caption{\label{fg:cut} Entanglement for N=100 as a function of a)
$\theta$ for $m_{0}=0$ and b) $m_{0}$ for $\theta=\pi/2$.}
\end{figure}

We plot in Fig.(\ref{fg:ent}a) the entanglement between the modes
for different ground states $m_{0}$ as a function of $\theta$ for
$N=1000$ particles. We find that the entanglement is generally
higher when $m_{0}$ is small. This corresponds to a regime where the
probability distribution of particles has many maximum peaks. This
occurs when the elastic collision rate becomes comparable to
the frequency difference between the modes in the absence of
Josephson-type interaction, $A_1\approx A_2$. But note that by
turning on the Josephson interaction with $\theta$ close to $\pi/2$,
there is a local minimum at $m_{0}=0$. In Fig. (\ref{fg:cut}b) we
plot entanglement at $\theta=\pi/2$ and observe this local minimum.
This means that for a large Josephson-type interaction strength,
there is a local minimum when $A_2>A_1$.

We have learned something interesting about the role of two-body
collisions in the generation of entanglement in a Bose-Einstein
condensate. Higher degrees of entanglement are generated by
switching the Josephson-type interaction when the rate of
collisions  is
comparable to the frequency difference between the modes (in the absence of
the Josepson-type interaction) $A_1\approx A_2$. However, if $A_2>A_1$, lower degrees of
entanglement are generated. In the case of a double well condensate,
if the collision probability in the absence of tunneling is
comparable (but lower) to the energy difference between the wells,
then, by letting the atoms tunnel lowering the barrier maximally,
entanglement will be maximum if the wells are symmetric. However, if
the collision probability exceeds the energy difference between the
wells in the absence of tunneling, maximum degrees of entanglement
are achieved when lowering maximally the barrier if the wells are
slightly asymmetric. This will favor the generation of entanglement.
In the case of a Bose-Einstein condensate consisting of two internal
degrees of freedom, if collisions occur with probability comparable
(but lower) to the energy difference between the modes, entanglement
will become maximum by coupling the internal degrees with a strong
resonant laser. But if the collision probability exceeds the energy
difference between the internal levels (in the absence of the
laser), it becomes more convenient for the generation of
entanglement, to turn on a slightly detuned laser coupling. The same
conclusions can we drawn from the analysis of Fig. (\ref{fg:cut}a)
where we plot the entanglement at $m_{0}=0$, i.e. $A_2>A_1$. We
observe local minimum at $\theta=\pi/2$. Therefore, we conclude that
in this case, it is more convenient to have $\theta$ sightly
different to $\pi/2$.

 In Fig.
 (\ref{fg:ent}b) we plot the entanglement as a function of the number of
 particles $N$ and different ground states $m_{0}$ for $\theta=\pi/2$. We can see that the
 entanglement is stronger as we have more particles in the
 condensate. We plotted the entanglement as a function of the number
 of particles for N as big as $10^{5}$ and found the same behavior.

 Therefore, to create high degrees of
 entanglement between the modes, it is convenient to have a condensate with large number of particles where the two-body elastic
 collision rate is comparable (but not larger) to the frequency difference between the modes in the absence of the Josephson-type interaction.
 One then turns on a
 strong near resonant laser (or lower maximally the tunneling barrier in an almost symmetric well).


\section{Discussion and outlook }
We have presented several predictions of a new model describing a
two-mode Bose-Einstein condensate which includes spin-exchange
inelastic collisions. These collisions have been observed in
experiments when they lead to particle loss. Nevertheless, there are
good reasons to believe that these processes do not always lead to
particle loss. The particles escape the trap after an inelastic
process only when there is an excess of energy larger than the
trapping potential. Our work is the first to analyze the effects of
inelastic collisions in the system when the particles do not escape
the trap. We show that such inelastic collisions produce important
qualitative and quantitative effects. Moreover, our model has the
great advantage of having exact analytical solutions for a set of
parameters. Previously, the study of the two-mode Bose-Einstein
condensate was restricted to numerical treatment.

Inelastic collisions which do not lead to particle loss have not
been yet a topic of experimental research. The main intension of our
work is to present a number of predictions in the evolution of the
relative population of the condensate and in the probability
distribution of particles in the system which would allow to detect
such inelastic process in the laboratory. The presence of these
effects can be easily detected by comparing the results to the
canonical two-mode hamiltonian which does not include inelastic
collisions. We find that the presence of inelastic collisions
produce multicomponent superposition states, rather than two
component cat states as predicted by the canonical model. We show
how inelastic processes have qualitative effects in evolution of the
relative population and are capable of preventing the self trapping
effect from occurring, even when the collision rate is larger than
the Josephson-type coupling. Taking advantage of the analytical
solution of the system, we include in our work a study on mode
entanglement and discuss under what conditions high degrees of
entanglement can be generated. We find that higher degrees of
entanglement can be generated when the condensate consists of large
number of particles and the two-body collision probability is
comparable (but not larger) to the energy difference between the
modes in absence of Josephson-type interaction. In such condensates,
maximally degrees of entanglement are generated by a strong resonant
laser coupling (or small symmetric tunneling barrier in the case of
a double well condensate). The analytical solution to our model
helped us learn details about the role of two-body collisions in the
generation of entanglement in the system. We found that collisions
favor entanglement as long as the collision probability does not
exceed the frequency difference between the modes (in the absence of
Josephson-type interaction). If such is the case, higher degrees of
entanglement can achieved by allowing maximal tunneling in a
slightly asymmetric well (or turning on a strong but slightly
detuned laser coupling).

To complete our study of inelastic spin-exchange collisions in the
two-mode Bose-Einstein condensate, we plan to introduce the loss of
particles due to inelastic collisions as a decoherence process using
a master equation. We are currently working in the generalization of
our model to spin-1 (or three well) condensates. We are also
studding the effects of many-body collisions in the system.

We thank C. G. Hern\'andez-Salinas for his help verifying some of
our results.

\end{document}